\documentclass[reprint,twocolumn,prl,showpacs,showkeys,aps]{revtex4-1}

\usepackage{epsfig,amsmath,amsfonts,amssymb}
\usepackage{amsmath,amsfonts,amssymb}
\usepackage{graphicx}
\usepackage{subfigure}
\usepackage{mathbbol}
\usepackage{amsthm}

\begin{document}

\title{Boosting Entanglement Between Exciton-Polaritons with On-Off Switching of Josephson Coupling}

\author{Dionisis Stefanatos}
\email{dionisis@post.harvard.edu}
\author{Emmanuel Paspalakis}%
\affiliation{Materials Science Department, School of Natural Sciences, University of Patras, Patras 26504, Greece}

\date{\today}

\begin{abstract}
We show that the appropriate on-off switching of Josephson coupling between exciton-polaritons in coupled semiconductor microcavities can reveal the full capacity for generating entanglement with a recently proposed method which essentially enhances the nonlinearity of the system. The improvement achieved with this simple modulation of the coupling is substantial over the case where it is kept constant. The suggested procedure is expected to find also application in other research areas, where nonlinear interacting bosons are encountered.
\end{abstract}

\maketitle

\emph{Introduction}.-Exciton-polaritons in semiconductor microcavities are hybrid light-matter quantum quasiparticles obeying Bose statistics, which emerge due to the strong coupling between cavity photons and excitons (electron-hole bound states) of the embedded semiconductor quantum well \cite{Lai07,Cerda10,Byrnes14,Whittaker18}. Because they can be easily created, controlled and detected, they have attracted considerable attention as candidates for the implementation of several critical quantum technologies. Applications include the generation of non-classical light \cite{Liew10,Adiyatullin17,Flayac17a,Flayac17b,Flayac17b,Klaas18}, qubits and gates for quantum computation \cite{Demirchyan14,Kyriienko16}, and solid state quantum simulators (special purpose quantum computers addressing complex problems) \cite{Kim17,Askitopoulos15,Ohadi17,Sigurdsson17,Berloff17,Kalinin17}.

In this context, a step of utmost importance is the creation of entanglement between exciton-polaritons in coupled cavities, since these systems of nonlinear interacting quantum oscillators can be exploited for quantum information processing with continuous variables \cite{Braunstein05}. The generation of entanglement relies on the strength of the nonlinearity, which is weak for semiconductor microcavities. Consequently, entanglement in these systems appears only as a perturbation \cite{Stefanatos18}, and this remains the case even at high densities, where the nonlinear effects become important, since the mean field approximation provides a fair classical description of the main system behavior \cite{Casteels17,Sun17}. In order to overcome this problem, a way to essentially amplify the nonlinearity strength in semiconductor microcavities using two coherent laser fields was recently suggested \cite{Liew18}, leading in theory to the creation of a fair amount of entanglement between exciton-polaritons in coupled cavities and networks in general. Throughout this process, the Josephson coupling between two coupled cavities is held constant.

In the present work, we provide analytical and numerical evidence that the entanglement generated with the procedure introduced in Ref.\ \cite{Liew18} can be substantially enhanced with the appropriate on-off switching of Josephson coupling between the cavities, exploiting thus the full capacity of the method. Since the model of two nonlinear interacting bosons is encountered in a wide spectrum of physical settings, we expect that the suggested methodology is not restricted only to exciton-polariton systems in semiconductor microcavities but can also find application in other contexts.

\emph{Model}.-We consider a pair of coupled cavities as in \cite{Liew18}, which can be implemented with the techniques of \cite{Vasconcelos11}, described by the Hamiltonian
\begin{equation}
\label{Hamiltonian}
H=\frac{\alpha}{2}(\hat{a}_1^2+\hat{a}_2^2+\hat{a}_1^{\dagger 2}+\hat{a}_2^{\dagger 2})-\mathcal{J}(t)(\hat{a}_1^{\dagger}\hat{a}_2+\hat{a}_1\hat{a}_2^{\dagger}).
\end{equation}
The first part of the Hamiltonian originates from an inverse four-wave mixing process in each cavity. As described in Ref.\ \cite{Liew18}, the starting point is the Hamiltonian $H_i=\alpha_0(\hat{a}_i^{\dagger}\hat{a}_i^{\dagger}\hat{a}_L\hat{a}_U+\hat{a}_L^\dagger\hat{a}_U^\dagger\hat{a}_i\hat{a}_i)/2$, where $\alpha_0$ is the strength of this nonlinear process, typically weak compared to the dissipation rate $\Gamma$ in optical systems. When the modes $\hat{a}_L,\hat{a}_U$ are driven by coherent laser fields, which can be described classically, pairs of particles scatter from $\hat{a}_L,\hat{a}_U$ to mode $\hat{a}_i$ and we are left with the first part of Hamiltonian (\ref{Hamiltonian}), where $\alpha=\alpha_0\langle a_L\rangle\langle a_U\rangle$
 is the nonlinearity enhanced by the classical field amplitudes, which can reach the regime $\alpha\gg\Gamma$. This is the advantage over the usual method where the central mode $\hat{a}_i$ is excited and correlations are created between $\hat{a}_L,\hat{a}_U$ \cite{Schwendimann03,Karr04R,Karr04,Savasta05,Portolan09,Romanelli10}, which requires a nonlinearity $\alpha_0$ stronger than the dissipation rate.
The second part is the familiar Josephson coupling, where $\mathcal{J}(t)$ is considered to be a function of time, restricted between zero and a maximum allowed value $0\leq \mathcal{J}(t)\leq J$, which can be controlled by external electric \cite{Christmann10} or optical \cite{Amo10} fields. We consider the situation where the coupling upper bound satisfies $J>\alpha>0$.

For a system of two oscillators coupled with a quadratic Hamiltonian like (\ref{Hamiltonian}) and starting from vacuum, the states are Gaussian and completely characterized by the second moments of the creation and annihilation operators of the two resonators, for example $\hat{a}_1^\dag\hat{a}_1, \hat{a}_2^\dag\hat{a}_2, \hat{a}_1^\dag\hat{a}_2, \hat{a}_1^2$ etc., while the first moments are zero due to the initial conditions. Instead of using directly the second moment operators, we can use specific linear combinations of them, a set of ten operators introduced by Dirac to describe exactly two coupled quantum oscillators \cite{Dirac63}, which are the generators of the symplectic group $Sp(4)$ \cite{Kim16,Stefanatos16,Stefanatos17}. Under the evolution described by Hamiltonian (\ref{Hamiltonian}), a closed set of differential equations can be obtained for the expectation values of these operators. Specifically, they are actually grouped into subsystems which are linear and homogenous in these variables. When starting from vacuum, only the subsystem formed by the following three operators
\begin{subequations}
\label{operators}
\begin{eqnarray}
\hat{S}_1&=&\frac{1}{2}(\hat{a}_1^\dagger\hat{a}_1+\hat{a}_2\hat{a}_2^\dag),\label{S1}\\
\hat{S}_2&=&\frac{i}{4}(\hat{a}_1^{\dagger 2}+\hat{a}_2^{\dagger 2}-\hat{a}_1^2-\hat{a}_2^2),\label{S2}\\
\hat{S}_3&=&\frac{1}{2}(\hat{a}_1^{\dagger}\hat{a}_2^{\dagger}+\hat{a}_1\hat{a}_2).\label{S3}
\end{eqnarray}
\end{subequations}
has nonzero initial conditions; the rest of the operators remain zero throughout and can be ignored.
Using Ehrenfest theorem for operators without explicit time dependence $d\langle \hat{A}\rangle/dt=\imath[H,\hat{A}]$ ($\hbar=1$), we find that the corresponding expectation values $S_i=\langle\hat{S}_i\rangle$, $i=1,2,3$ satisfy the following system of equations
\begin{subequations}
\label{system}
\begin{eqnarray}
\dot{S}_1&=&-2\alpha S_2,\label{system1}\\
\dot{S}_2&=&-2\alpha S_1+2\mathcal{J}S_3,\label{system2}\\
\dot{S}_3&=&-2\mathcal{J}S_2,\label{system3}
\end{eqnarray}
\end{subequations}
with initial conditions $S_1(0)=1/2, S_2(0)=S_3(0)=0$.
Under the above evolution, the following constant of the motion can be easily verified
\begin{equation}
\label{constant}
S_1^2-S_2^2-S_3^2=1/4.
\end{equation}

\emph{Entanglement quantification}.-We will characterize the entanglement between the two coupled oscillators using the covariance matrix $V$ of the corresponding position and momentum operators $\hat{q}_i, \hat{p}_i, i=1,2$. If we define $(\hat{x}_1,\hat{x}_2,\hat{x}_3,\hat{x}_4)=(\hat{q}_1,\hat{p}_1,\hat{q}_2,\hat{p}_2)$, then the elements of the corresponding covariance matrix become $V_{ij}=\langle\hat{x}_i\hat{x}_j+\hat{x}_j\hat{x}_i\rangle/2$, where note that the first moments are zero due to the vacuum initial conditions. These elements can be expressed in terms of the second moments of creation and annihilation operators and consequently the nonzero values $S_i$ as
\begin{equation*}
V=
\left(
\begin{array}{cc}
  A & C \\
  C^T & B
\end{array}
\right)=
\left(
\begin{array}{cccc}
  S_1 & S_2 & S_3 & 0 \\
  S_2 & S_1 & 0 & -S_3 \\
  S_3 & 0 & S_1 & S_2 \\
  0 & -S_3 & S_2 & S_1
\end{array}
\right).
\end{equation*}
In order to quantify entanglement we will use the logarithmic negativity \cite{Vidal02,Eisert03,Plenio05}, a quantity which for two-mode Gaussian states actually measures the squeezing of appropriate field quadratures \cite{Duan00,Simon00}. For this particular case the logarithmic negativity is given by $\mathcal{N}=\mbox{max}[0, -\ln(2\tilde{\nu}_-)]$, where $\tilde{\nu}_-$ is the smallest symplectic eigenvalue of a modified covariance matrix $\tilde{V}$ corresponding to the partially transposed state. We can evaluate $\tilde{\nu}_-$ in terms of $S_i$ using the formula \cite{Adesso04}
\begin{equation*}
\tilde{\nu}_-=\sqrt{\frac{\tilde{\Delta}(V)-\sqrt{\tilde{\Delta}^2(V)-4\mbox{det}V}}{2}},
\end{equation*}
where $\tilde{\Delta}(V)=\mbox{det}A+\mbox{det}B-2\mbox{det}C=2(S_1^2-S_2^2+S_3^2)$ and $\mbox{det}V=(S_1^2-S_2^2-S_3^2)^2$ ,
from which we obtain
\begin{equation*}
\tilde{\nu}_-=\sqrt{S_1^2-S_2^2}-|S_3|<1/2.
\end{equation*}
The last inequality can be proved using (\ref{constant}), and the logarithmic negativity is given by the expression
\begin{equation}
\label{negativity}
\mathcal{N}=-\ln(2\tilde{\nu}_-)=-\ln\left[2\left(\sqrt{S_1^2-S_2^2}-|S_3|\right)\right].
\end{equation}

\begin{figure}[t!]
\centering
\subfigure[\ Josephson couplings]{
\label{fig:controls}
\includegraphics[scale=0.28]{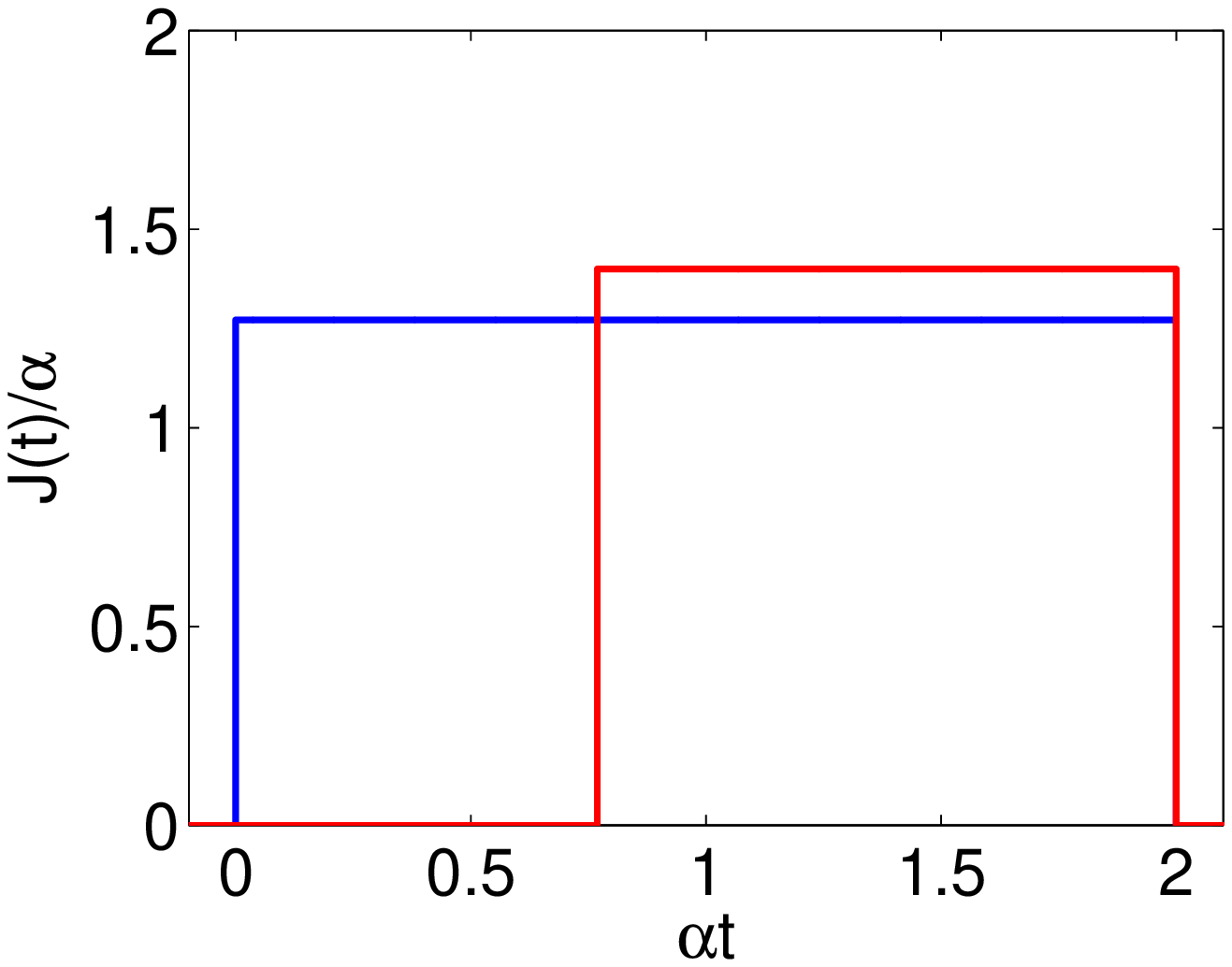}}
\subfigure[\ Corresponding trajectories]{
\label{fig:trajectories}
\includegraphics[scale=0.28]{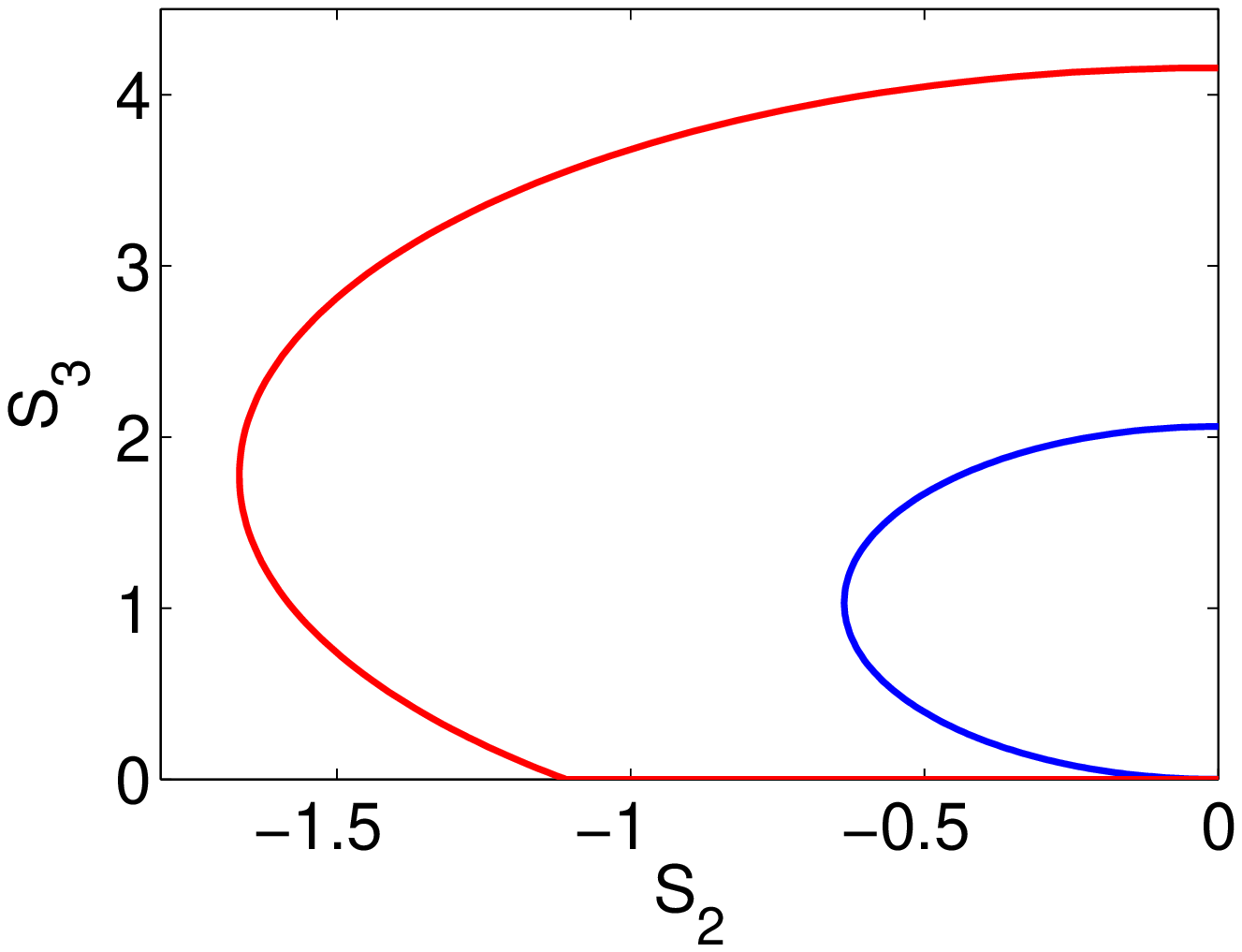}}
\caption{(Color online) (a) Constant Josephson coupling (blue line) and modified coupling (\ref{control}) (red line), for a common duration $\alpha T=2$. The amplitude of the constant pulse is found from (\ref{u_T}) to be $u_T=J_T/\alpha=1.2716$, while for the second (nonzero) pulse in the modified coupling it is $u=J/\alpha=1.4$ (b) Corresponding trajectories on the $S_2S_3$ plane, where the inner blue trajectory correspond to the constant coupling and the outer red trajectory to the modified coupling.}
\label{fig:controls_trajectories}
\end{figure}

\begin{figure}[t]
\centering
\subfigure[\ No dissipation $\Gamma=0$]{
\label{fig:nodissipation}
\includegraphics[scale=0.28]{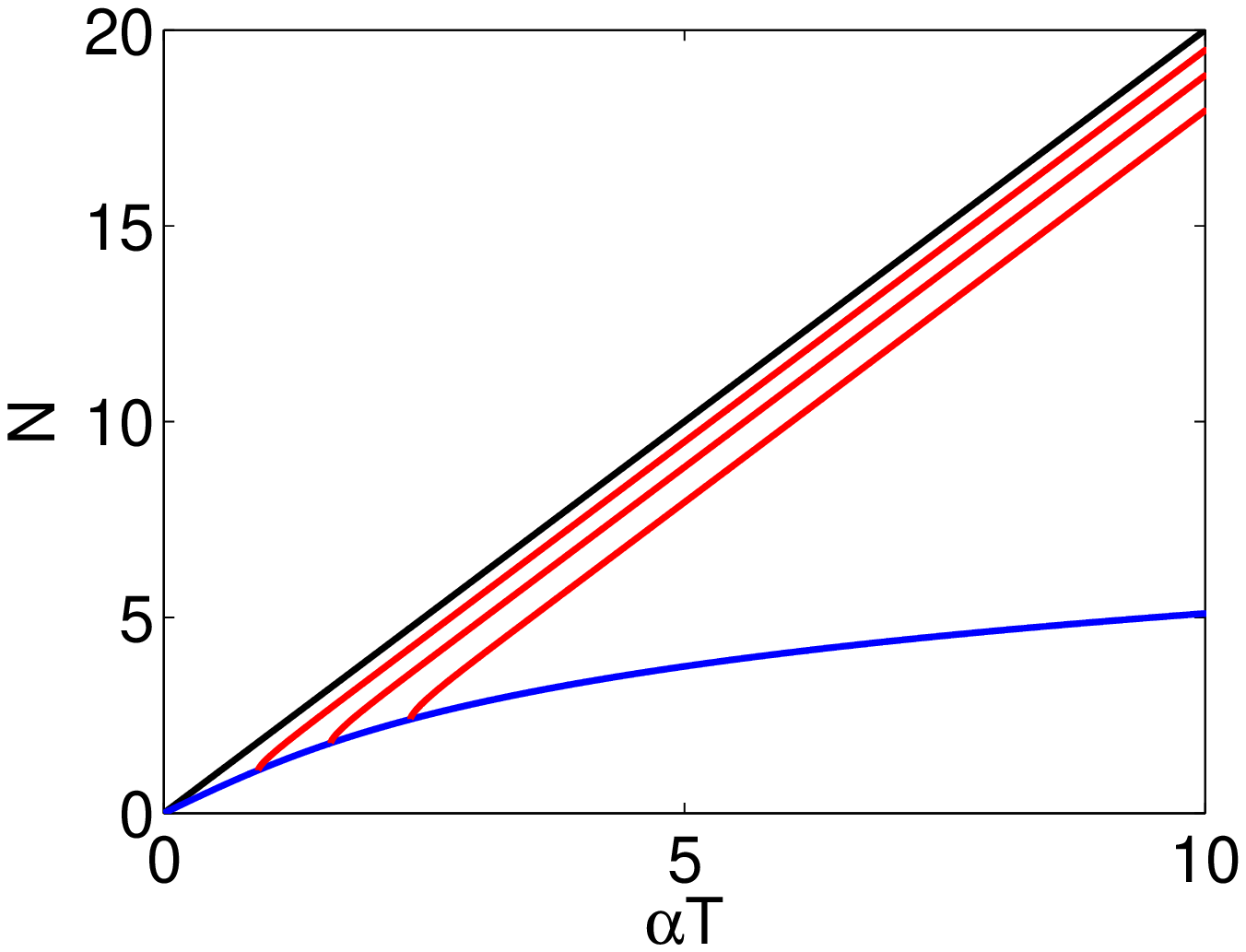}}
\subfigure[\ Dissipation $\Gamma=0.2\alpha$]{
\label{fig:dissipation}
\includegraphics[scale=0.28]{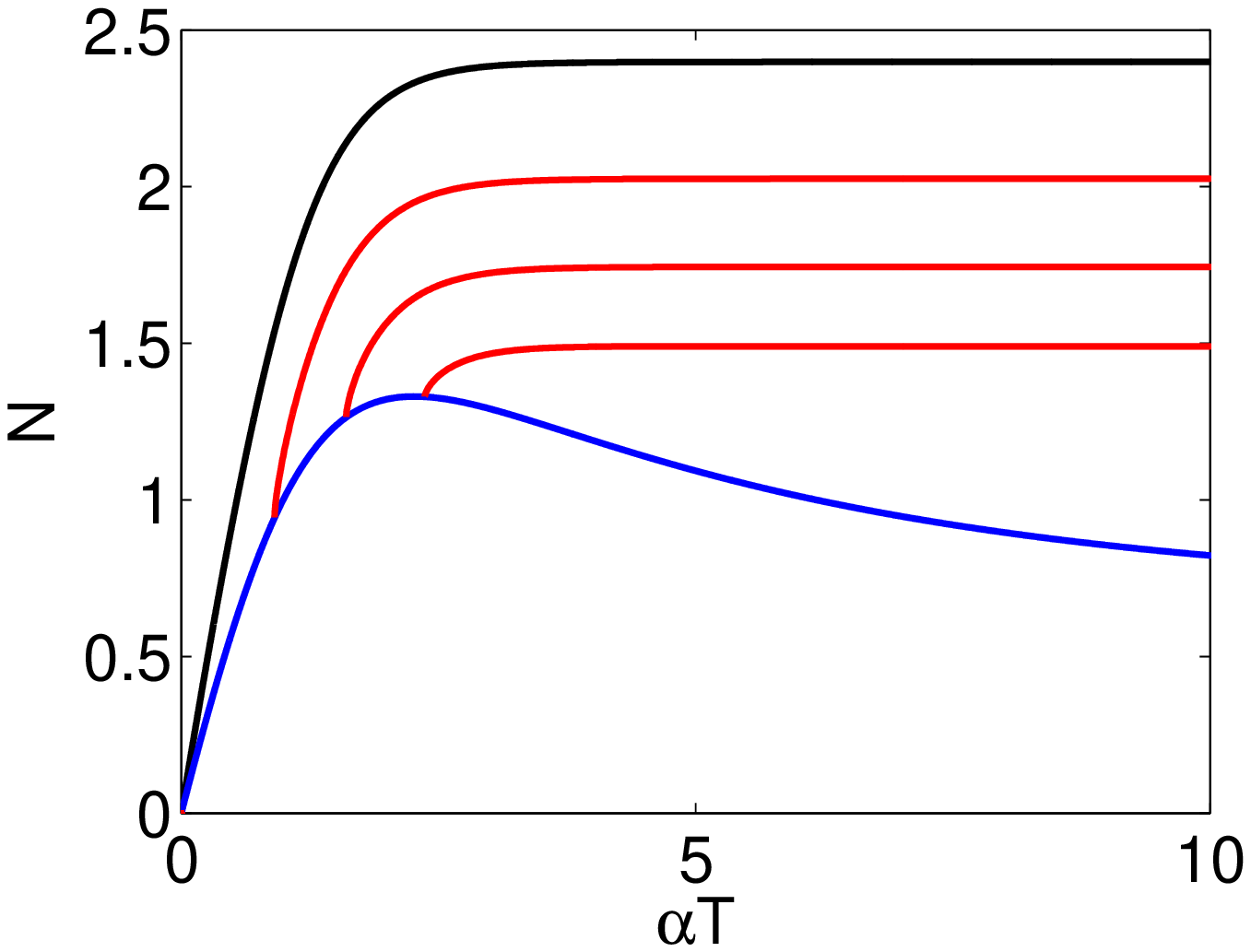}}
\caption{(Color online) Logarithmic negativity as a function of duration $\alpha T$ (a) In the absence of dissipation, $\Gamma=0$. The lower blue curve corresponds to the constant coupling with amplitude $u_T=J_T/\alpha$ given in (\ref{u_T}), while the upper black curve corresponds to the modified coupling (\ref{control}) with a delta final pulse (infinite $u=J/\alpha$). The intermediate red lines correspond to finite values of $u=J/\alpha=1.2,1.4,2$ (from bottom to top) (b) Same as in (a) but in the presence of dissipation $\Gamma=0.2\alpha$.}
\label{fig:negativity}
\end{figure}

\emph{Constant Josephson coupling}.-The authors of Ref.\ \cite{Liew18} consider the situation where a constant coupling $\mathcal{J}(t)=J_T$, with $\alpha<J_T\leq J$, is applied for the whole time interval $0\leq t\leq T$. The appropriate value of the coupling depends on $T$, as it is denoted by the subscript in $J_T$.
By taking the time derivative of (\ref{system2}) and using (\ref{system1}), (\ref{system3}) we obtain the following differential equation for $S_2$
\begin{equation}
\label{Diff_S2}
\ddot{S}_2+4(J_T^2-\alpha^2)S_2=0.
\end{equation}
Solving for the initial conditions $S_2(0)=0, \dot{S}_2(0)=-\alpha$ we find
\begin{equation*}
S_2(t)=-\frac{\sin(2\omega\alpha t)}{2\omega},\quad S_3(t)=\frac{u_T[1-\cos(2\omega\alpha t)]}{2(u_T^2-1)},
\end{equation*}
where the normalized angular frequency is $\omega=\sqrt{u_T^2-1}$ and $u_T=J_T/\alpha$.
The choice
\begin{equation}
\label{u_T}
2\omega\alpha T=\pi\Rightarrow u_T=\frac{J_T}{\alpha}=\sqrt{1+\left(\frac{\pi}{2\alpha T}\right)^2} ,
\end{equation}
leads to $S_2(T)=0, S_3(T)=u_T/(u_T^2-1)$. From the constant of the motion (\ref{constant}) we can also obtain $S_1(T)=\sqrt{S_2^2(T)+S_3^2(T)+1/4}$, where note that $S_1(T)>0$ from (\ref{S1}). Putting these values in (\ref{negativity}), we finally find the logarithmic negativity as a function of the final time $T$
\begin{equation}
\label{old_negativity}
\mathcal{N}_0=2\ln\left\{\frac{2\alpha T}{\pi}\left[\sqrt{1+\left(\frac{\pi}{2\alpha T}\right)^2}+1\right]\right\},
\end{equation}
where the subscript denotes the absence of dissipation. Observe that in the limit of large $T$ the logarithmic negativity increases logarithmically with time. A constant pulse of duration $\alpha T=2$ is shown in Fig. \ref{fig:controls}, with amplitude $u_T=J_T/\alpha=1.2716$, as calculated from (\ref{u_T}). The corresponding trajectory on the $S_2S_3$ plane is displayed in Fig. \ref{fig:trajectories} (inner blue line). In Fig. \ref{fig:nodissipation} we plot the logarithmic negativity (\ref{old_negativity}) as a function of the duration $T$ (lower blue line).

\emph{On-off switching}.-We show that by the appropriate on-off switching of the Josephson coupling and using the maximum available value $J$, a substantial improvement of the entanglement generated within the same duration $T$ can be obtained. We first describe the motivation behind this modified coupling and then derive the corresponding performance. Using the constant of the motion (\ref{constant}), the expression (\ref{negativity}) for the logarithmic negativity becomes
\begin{equation}
\label{S3_negativity}
\mathcal{N}=\ln\left[2\left(\sqrt{S_3^2+1/4}+|S_3|\right)\right],
\end{equation}
which is an increasing function of $|S_3|$. System (\ref{system}) can be reduced to a two-dimensional system for $S_2, S_3$ only and, if we use polar coordinates on the $S_2S_3$ plane, defined as $\rho=\sqrt{S_2^2+S_3^2}$, $\tan(\pi-\phi)=-\tan{\phi}=S_3/S_2$ (angle $\phi$ is measured from the negative $S_2$-axis), we obtain the equations
\begin{subequations}
\label{polar}
\begin{eqnarray}
\dot{\rho}&=&2\alpha\cos{\phi}\sqrt{\rho^2+1/4}, \label{ro}\\
\dot{\phi}&=&2\mathcal{J}-2\alpha\frac{\sin{\phi}}{\rho}\sqrt{\rho^2+1/4},\label{fi}
\end{eqnarray}
\end{subequations}
with initial conditions $\rho(0)=0, \phi(0)=0$. Note that the initial value of $\phi$ is determined not only from the initial values $S_i(0)$ but also from Eqs. (\ref{system2}), (\ref{system3}) which, for $t=0^+$ give $S_2(0^+)<0, S_3(0^+)=0$, thus $\phi(0)=0$. If we had chosen the usual definition for $\phi$ ($\tan{\phi}=S_3/S_2$), then the initial value would be $\pi$ instead of $0$. In order to maximize the logarithmic negativity (\ref{S3_negativity}) at the final time $t=T$, we simply need to maximize $S_3(T)=\rho(T)\sin{\phi(T)}$. We set $\phi(T)=\pi/2$ and require the maximization of the final value $\rho(T)$. But from (\ref{ro}) observe that $\dot{\rho}$ is maximized when $\phi=0$ $(\cos{\phi}=1)$. This crucial observation provides the motivation for the suggested modified coupling. If the upper bound $J$ of the Josephson coupling was infinite, then the optimal strategy would be the following: turn off the coupling $\mathcal{J}(t)=0$ for the time interval $[0\,T)$ in order to build the maximum $\rho(T)$, and at the final time $t=T$ apply a delta pulse to rotate this maximum value instantaneously to $S_3(T)$.
When $J$ is finite the duration $t_2$ of the final pulse is nonzero, the duration of the initial zero pulse becomes $t_1=T-t_2$, and the applied modified coupling consists of the pulse sequence
\begin{equation}
\label{control}
\mathcal{J}(t)=\left\{\begin{array}{cl} 0, & 0\leq t\leq t_1 \\J, & t_1<t\leq T=t_1+t_2\end{array}\right..
\end{equation}
In Fig. \ref{fig:controls} we display this modified coupling with duration $\alpha T=2$ and maximum amplitude $u=J/\alpha=1.4$ (red curve), while in Fig. \ref{fig:trajectories} we plot the corresponding trajectory (outer red curve). The first part of the trajectory, along $\phi=0$, corresponds to a \emph{singular arc} in optimal control terminology \cite{Lapert10}. Observe that with an amplitude slightly higher than that of the constant coupling with the same duration ($u_T=J_T/\alpha=1.2716$), a much larger final value $S_3(T)$, and thus of the negativity, is achieved.

Having motivated the superiority of modified coupling (\ref{control}) over the simple constant coupling, we next move to calculate the switching time $t_1$ and the final logarithmic negativity, when the total duration $T$ is given. In the interval $0\leq t\leq t_1$, where $\mathcal{J}(t)=0$, Eq. (\ref{Diff_S2}) becomes $\ddot{S}_2-4\alpha^2S_2=0$ and we easily obtain $S_2(t_1)=-\sinh(2\alpha t_1)/2$, while from (\ref{system3}) we have $S_3(t_1)=S_3(0)=0$. During the subsequent interval $t_1<t\leq t_1+t_2$ it is $\mathcal{J}(t)=J$ and $S_2$ satisfies (\ref{Diff_S2}) with $J_T$ replaced by $J$. Solving for the initial conditions at $t=t_1$ and using also (\ref{system3}) we find at the final time $t=t_1+t_2=T$
\begin{eqnarray}
S_2(T)&=&-\frac{1}{2}\sinh(2\alpha t_1)\cos(2\omega\alpha t_2)\nonumber\\
 &&-\frac{1}{2\omega}\cosh(2\alpha t_1)\sin(2\omega\alpha t_2) , \label{S2_final}
\end{eqnarray}
and
\begin{eqnarray}
S_3(T)&=&\frac{u}{2\omega}\sinh(2\alpha t_1)\sin(2\omega\alpha t_2)\nonumber\\
 &&+\frac{u}{2\omega^2}\cosh(2\alpha t_1)[1-\cos(2\omega\alpha t_2)] \label{S3_final},
\end{eqnarray}
where now $\omega=\sqrt{u^2-1}$ and $u=J/\alpha$.
The choice $\tan(2\omega\alpha t_2)=-\omega\tanh(2\alpha t_1)$ gives $S_2(T)=0$ while at the same time maximizes $S_3(T)$. We can use this relation to eliminate $t_2$ and obtain the following transcendental equation for the duration $t_1$ of the zero pulse
\begin{equation}
\label{transcendental}
T=t_1+t_2=t_1+\frac{\pi-\tan^{-1}[\omega\tanh(2\alpha t_1)]}{2\omega\alpha},
\end{equation}
where $T$ is the total duration. The duration $t_2$ of the second pulse can be easily determined after we find $t_1$. The left hand side of (\ref{transcendental}) is an increasing function of $t_1$, thus this equation has at most one solution. Setting $t_1=0$ we find the minimum necessary duration $T>\pi/(2\omega\alpha)$ such that (\ref{transcendental}) has a (unique) solution for a specific $J$. Using this relation the other way around we obtain that for a specific duration $T$ the transcendental equation has a (unique) solution for every $J$ greater than a T-dependent threshold
\begin{equation}
\label{condition}
u=\frac{J}{\alpha}>u_T=\sqrt{1+\left(\frac{\pi}{2\alpha T}\right)^2}.
\end{equation}

If we calculate $S_3(T)$ by eliminating $t_2$ in (\ref{S3_final}) and then plug this value in (\ref{S3_negativity}), we obtain the logarithmic negativity at the final time $t=T$ as a function of $t_1$ (which is an implicit function of $T$)
\begin{equation}
\label{new_negativity}
\mathcal{N}_{0,u}=\ln\left[\frac{u\cosh(2\alpha t_1)+\sqrt{1+u^2\sinh^2(2\alpha t_1)}}{u-1}\right].
\end{equation}
The first subscript denotes the absence of dissipation, as before, while the second one the dependence on $u$.
In the large time limit $at_1\gg 1$ we find $\mathcal{N}_{0,u}\rightarrow ue^{2\alpha t_1}/(u-1)$. But at the same limit $\tanh(2\alpha t_1)\rightarrow 1$, so $T\rightarrow t_1+t_2^{\infty}$, where
\begin{equation}
\label{t2_limit}
t_2^{\infty}=\frac{\pi-\tan^{-1}\omega}{2\omega\alpha} ,
\end{equation}
is a finite value. The large time limit of the logarithmic negativity becomes thus
\begin{equation}
\label{negativity_limit}
\mathcal{N}_{0,u}\rightarrow\ln\left(\frac{ue^{-2\alpha t_2^{\infty}}}{u-1}\right)+2\alpha T,
\end{equation}
which is a linear function of time, corresponding to an exponential squeezing. Note that the constant term (independent of $T$) in (\ref{negativity_limit}) is an increasing function of $u$. In the limit $u\gg 1$ ($J\gg\alpha$) it is $t_2^{\infty}\rightarrow 0$ and this negative term tends to zero. We thus obtain the ultimate bound on the entanglement which can be produced with the physical setting considered here
\begin{equation}
\label{ultimate}
\mathcal{N}_{0,\infty}=2\alpha T.
\end{equation}
In Fig. \ref{fig:nodissipation} we display this bound as a function of the duration (upper black line passing through the origin), while recall that the lower blue curve corresponds to the final negativity obtained with a constant coupling of the same duration. The intermediate red lines display the final logarithmic negativity (\ref{new_negativity}) obtained with the modified coupling (\ref{control}) for finite values of $u=J/\alpha=1.2,1.4,2$ (from bottom to top). Observe that each of these lines starts at a different duration $T$, as determined from (\ref{condition}) for each $u$ such that the corresponding transcendental equation has solution, while they approach the ultimate bound as $u$ increases.

\emph{Effect of dissipation}.-We consider the simple dissipation model of Ref.\ \cite{Liew18}, where each second order correlation dissipates at a rate $\Gamma$. The system equations are modified as
\begin{subequations}
\label{system_dissipation}
\begin{eqnarray}
\dot{S}_1&=&-\Gamma S_1-2\alpha S_2+\Gamma/2,\label{d_system1}\\
\dot{S}_2&=&-2\alpha S_1-\Gamma S_2+2\mathcal{J}S_3,\label{d_system2}\\
\dot{S}_3&=&-2\mathcal{J}S_2-\Gamma S_3,\label{d_system3}
\end{eqnarray}
\end{subequations}
where note the differentiation of $S_1=\langle\hat{a}_1^\dagger\hat{a}_1+\hat{a}_2\hat{a}_2^\dag\rangle/2=\langle\hat{a}_1^\dagger\hat{a}_1+\hat{a}_2^\dag\hat{a}_2+1\rangle/2$. If we define $\tilde{S}_i=e^{\Gamma t}S_i$ we obtain the system
\begin{eqnarray*}
\dot{\tilde{S}}_1&=&-2\alpha \tilde{S}_2+\Gamma e^{\Gamma t}/2,\\
\dot{\tilde{S}}_2&=&-2\alpha \tilde{S}_1+2\mathcal{J}\tilde{S}_3,\\
\dot{\tilde{S}}_3&=&-2\mathcal{J}\tilde{S}_2.
\end{eqnarray*}
For constant $\mathcal{J}(t)=J$ we can obtain a differential equation for $\tilde{S}_2$ analogous to Eq. (\ref{Diff_S2})
\begin{equation}
\label{Diff_diss_S2}
\ddot{\tilde{S}}_2+4(J^2-\alpha^2)\tilde{S}_2=-\alpha\Gamma e^{\Gamma t}.
\end{equation}
This equation can be easily solved analytically; once we have obtained $S_2(t)=e^{-\Gamma t}\tilde{S}_2(t)$ we can easily integrate (\ref{d_system1}), (\ref{d_system2}) and find
\begin{eqnarray*}
S_1(t)&=&e^{-\Gamma t}\left\{S_1(0)+\int_0^te^{\Gamma t'}[\Gamma/2-2\alpha S_2(t')]dt'\right\},\\
S_3(t)&=&e^{-\Gamma t}\left[S_3(0)-2J\int_0^te^{\Gamma t'}S_2(t')dt'\right].
\end{eqnarray*}

Using the above formulas, we have been able to obtain analytical results for the final values $S_i(T)$, extensively verified with numerical simulations of system (\ref{system_dissipation}), for both the previously presented strategies, the constant coupling $\mathcal{J}(t)=J_T$ and the modified coupling (\ref{control}). The corresponding expressions are cumbersome and are not displayed here, but we have used them to create plots of the logarithmic negativity as a function of duration $T$, displayed in Fig. \ref{fig:dissipation}. Observe that, in the presence of dissipation, the logarithmic negativity attained with the modified coupling is saturated for large durations. We have calculated analytically this saturation limit as
\begin{eqnarray}
\label{new_limit}
\mathcal{N}_{\gamma,u}&\rightarrow&-\ln\Biggl\{1-\frac{1+\gamma/u}{1+\gamma}e^{-\Gamma t_2^{\infty}}+\frac{1}{\gamma^2+u^2-1}\times\\
&&\left\{\left[\frac{\gamma^2}{u}+(u-1)\left(1+\frac{\gamma}{u}\right)\right]e^{-\Gamma t_2^{\infty}}-(u-1)\right\}\Biggr\},\nonumber
\end{eqnarray}
where $\gamma=\Gamma/(2\alpha)$, $u=J/\alpha$ and $t_2^{\infty}$ is given in (\ref{t2_limit}).
The ultimate bound in the presence of dissipation (limiting value of the upper black curve), for large times $\alpha T\gg 1$ and $u\gg 1$ ($J\gg\alpha$), is
\begin{equation}
\label{ultimate_dissipation}
\mathcal{N}_{\gamma,\infty}\rightarrow\ln\left(1+\frac{1}{\gamma}\right).
\end{equation}
The logarithmic negativity obtained with the constant coupling strategy (lower blue curve) attains a maximum value, while for large times it also converges to a constant value
\begin{equation}
\label{old_limit}
\mathcal{N}_{\gamma}\rightarrow-\ln\left(\frac{\sqrt{\gamma^4+\gamma^2+1}-1}{\gamma^2}\right).
\end{equation}

\emph{Conclusion}.
We have shown that, with the appropriate modulation of Josephson coupling between two exciton-polariton cavities, the entanglement generated by a recently proposed method which effectively amplifies the system nonlinearity can be substantially enhanced. This work can find immediate application in quantum information processing with polaritons, but also in other areas where nonlinear interacting bosons are encountered. The presented results can be further improved in the presence of dissipation by using optimal control methods \cite{Lapert10,Goerz15}.

\emph{Acknowledgements}. Co-financed by Greece and the European Union - European Regional Development Fund via the General Secretariat for Research and Technology bilateral Greek-Russian Science and Technology collaboration project on Quantum Technologies (project code name POLISIMULATOR).


\begin{thebibliography}{99}

\bibitem{Lai07}
C.W. Lai, N.Y. Kim, S. Utsunomiya, G. Roumpos, H. Deng, M.D. Fraser, T. Byrnes, P. Recher, N. Kumada, T. Fujisawa, and Y. Yamamoto
Nature 450, 529 (2007).

\bibitem{Cerda10}
E.A. Cerda-Méndez, D.N. Krizhanovskii, M. Wouters, R. Bradley, K. Biermann, K. Guda, R. Hey, P.V. Santos, D. Sarkar, and M.S. Skolnick, Phys. Rev. Lett. 105, 116402 (2010).

\bibitem{Byrnes14}
T. Byrnes, N.Y. Kim, and Y. Yamamoto, Nat. Phys. 10, 803 (2014).

\bibitem{Whittaker18}
C.E. Whittaker, E. Cancellieri, P.M. Walker, D.R. Gulevich, H. Schomerus, D. Vaitiekus, B. Royall, D.M. Whittaker, E. Clarke, I.V. Iorsh, I.A. Shelykh, M.S. Skolnick, and D.N. Krizhanovskii, Phys. Rev. Lett. 120, 097401 (2018).


\bibitem{Liew10}
T.C.H. Liew and V. Savona, Phys. Rev. Lett. 104, 183601 (2010).

\bibitem{Adiyatullin17}
A.F. Adiyatullin, M.D. Anderson, H. Flayac, M.T. Portella-Oberli, F. Jabeen,
C. Ouellet-Plamondon, G.C. Sallen, and B. Deveaud, Nat. Commun. 8, 1329 (2017).

\bibitem{Flayac17a}
H. Flayac and V. Savona, Phys. Rev. A 95, 043838 (2017).

\bibitem{Flayac17b}
H. Flayac and V. Savona
Phys. Rev. A 96, 053810 (2017).

\bibitem{Klaas18}
M. Klaas, H. Flayac, M. Amthor, I.G. Savenko, S. Brodbeck, T. Ala-Nissila, S. Klembt, C. Schneider, and S. H\"{o}fling, Phys. Rev. Lett. 120, 017401 (2018).

\bibitem{Demirchyan14}
S.S. Demirchyan, I.Yu. Chestnov, A.P. Alodjants, M.M. Glazov, and A.V. Kavokin, Phys. Rev. Lett. 112, 196403 (2014).


\bibitem{Kyriienko16}
O. Kyriienko and T.C.H. Liew, Phys. Rev. B 93, 035301 (2016)

\bibitem{Kim17}
N.Y. Kim and Y. Yamamoto in Quantum Simulations with Photons and Polaritons, edited by D. G. Angelakis, Quantum Science and Technology (Springer, Cham, 2017).
	
\bibitem{Askitopoulos15}
A. Askitopoulos, T.C.H. Liew, H. Ohadi, Z. Hatzopoulos, P.G. Savvidis, P.G. Lagoudakis, Phys. Rev. B 92, 035305 (2015).
	
\bibitem{Ohadi17}
H. Ohadi, A.J. Ramsay, H. Sigurdsson, Y. del Valle-Inclan Redondo, S.I. Tsintzos, Z. Hatzopoulos, T.C.H. Liew, I.A. Shelykh, Y.G. Rubo, P.G. Savvidis, and J.J. Baumberg, Phys. Rev. Lett. 119, 067401 (2017).

\bibitem{Sigurdsson17}
H. Sigurdsson, A.J. Ramsay, H. Ohadi, Y.G. Rubo, T.C.H. Liew, J.J. Baumberg, and I.A. Shelykh, Phys. Rev. B 96, 155403 (2017).

\bibitem{Berloff17}
N.G. Berloff, M. Silva, K. Kalinin, A. Askitopoulos, J.D. T\"{o}pfer, P. Cilibrizzi, W. Langbein, P.G. Lagoudakis, Nat. Mater. 16, 1120 (2017).

\bibitem{Kalinin17}
K.Kalinin, P.G. Lagoudakis, and N.G. Berloff, arXiv:1709.04683 (2017).

\bibitem{Braunstein05}
S.L. Braunstein and P. van Loock, Rev. Mod. Phys. 77, 513 (2005).


\bibitem{Stefanatos18}
D. Stefanatos and E. Paspalakis,  arXiv:1802.07097 (2018).

\bibitem{Casteels17}
W. Casteels and C. Ciuti, Phys. Rev. A 95, 013812 (2017).

\bibitem{Sun17}
M. Sun, I.G. Savenko, H. Flayac, and T.C.H. Liew, Sci. Rep. 7, 45243 (2017).

\bibitem{Liew18}
T.C.H. Liew and Y.G. Rubo, Phys. Rev. B 97, 041302(R) (2018).

\bibitem{Vasconcelos11}
S.M. de Vasconcellos, A. Calvar, A Dousse, J. Suffczy\'{n}ski, N. Dupuis, A. Lema\^{i}tre, I. Sagnes, J Bloch, P. Voisin, and P. Senellart, Appl. Phys. Lett. 99, 101103 (2011).

\bibitem{Schwendimann03}
P. Schwendimann, C. Ciuti, and A. Quattropani, Phys. Rev. B 68, 165324 (2003).

\bibitem{Karr04R}
J.Ph. Karr, A. Baas, R. Houdr\'{e}, and E. Giacobino, Phys. Rev. A 69, 031802(R) (2004).

\bibitem{Karr04}
J.Ph. Karr, A. Baas, and E. Giacobino, Phys. Rev. A 69, 063807 (2004).

\bibitem{Savasta05}
S. Savasta, O. Di Stefano, V. Savona, and W. Langbein
Phys. Rev. Lett. 94, 246401 (2005).

\bibitem{Portolan09}
S. Portolan, O. Di Stefano, S. Savasta, and V. Savona, EPL 88, 20003 (2009)

\bibitem{Romanelli10}
M. Romanelli, J. Ph. Karr, C. Leyder, E. Giacobino, and A. Bramati, Phys. Rev. B 82, 155313 (2010).

\bibitem{Christmann10}
G. Christmann, C. Coulson, J.J. Baumberg, N.T. Pelekanos, Z. Hatzopoulos, S.I. Tsintzos, and P.G. Savvidis, Phys. Rev. B 82, 113308 (2010).

\bibitem{Amo10}
A. Amo, S. Pigeon, C. Adrados, R. Houdre, E. Giacobino, C. Ciuti, and A. Bramati, Phys. Rev. B 82, 081301(R) (2010).

\bibitem{Dirac63}
P.A.M. Dirac, J. Math. Phys. 4, 901 (1963).

\bibitem{Kim16}
S. Ba\c{s}kal, Y.S. Kim, and M.E. Noz, Symmetry 8, 55 (2016).

\bibitem{Stefanatos16}
D. Stefanatos, Automatica 73, 71 (2016).

\bibitem{Stefanatos17}
D. Stefanatos, Quantum Sci. Technol. 2, 014003 (2017).

\bibitem{Vidal02}
G. Vidal and R.F. Werner, Phys. Rev. A 65, 032314 (2002).

\bibitem{Eisert03}
K. Audenaert, M.B. Plenio, and J. Eisert, Phys. Rev. Lett. 90, 027901 (2003).

\bibitem{Plenio05}
M.B. Plenio, Phys. Rev. Lett. 95, 090503 (2005).

\bibitem{Duan00}
L.-M. Duan, G. Giedke, J.I. Cirac, and P. Zoller, Phys. Rev. Lett. 84, 2722 (2000).

\bibitem{Simon00}
R. Simon, Phys. Rev. Lett. 84, 2726 (2000).

\bibitem{Adesso04}
G. Adesso, A. Serafini, and F. Illuminati, Phys. Rev. Lett. 92, 087901 (2004).

\bibitem{Lapert10}
M. Lapert, Y. Zhang, M. Braun, S.J. Glaser, and D. Sugny, Phys. Rev. Lett. 104, 083001 (2010).

\bibitem{Goerz15}
M.H. Goerz, G. Gualdi, D.M. Reich, C.P. Koch, F. Motzoi, K.B. Whaley, J. Vala, M.M. Muller, S. Montangero, and T. Calarco, Phys. Rev. A 91, 062307 (2015).







\end{thebibliography}
\end{document}